\documentclass[aps,prl,reprint,superscriptaddress,longbibliography,floatfix]{revtex4-2}
\usepackage{amsmath,amssymb,bm,graphicx}
\usepackage[colorlinks=true,citecolor=blue,urlcolor=blue,linkcolor=blue]{hyperref}
\usepackage[T1]{fontenc}
\usepackage{orcidlink}

\hypersetup{pdftitle={Non-Gaussian bosonic resources across phase transitions},
 pdfauthor={Paolo Stornati, Piotr Sierant, Emanuele Tirrito}}
\newcommand{\Q}{\mathcal Q}
\newcommand{\Tr}{\operatorname{Tr}}
\newcommand{\StringSpreadLow}{2.35}
\newcommand{\StringSpreadHigh}{2.46}
\newcommand{\StringFitOne}{2.42597}
\newcommand{\StringFitTwo}{2.44866}
\newcommand{\QREFitOne}{2.47021}
\newcommand{\QREFitTwo}{2.47039}
\newcommand{\MIQRECenter}{2.471}

\begin{document}
\title{Bosonic Magic Resources at Quantum Criticality}
\author{Paolo Stornati~\orcidlink{0000-0003-4708-9340}}
\affiliation{Barcelona Supercomputing Center (BSC), Barcelona, Spain}
\author{Piotr Sierant~\orcidlink{0000-0001-9219-7274}}
\affiliation{Barcelona Supercomputing Center (BSC), Barcelona, Spain}
\author{Emanuele Tirrito\orcidlink{0000-0001-7067-1203}}
\affiliation{Institute of Physics, École Polytechnique Fédérale de Lausanne (EPFL), CH-1015 Lausanne, Switzerland \\
Center for Quantum Science and Engineering, Ecole Polytechnique Fédérale de Lausanne (EPFL), CH-1015 Lausanne, Switzerland}
\begin{abstract}
Non-Gaussianity is a resource for bosonic quantum information processing, providing a necessary ingredient for universal quantum computation in continuous-variable systems.
Here we quantify non-Gaussianity in the ground state of the one-dimensional extended Bose–Hubbard model using quantum relative entropy (QRE), a measure determined solely by one-body correlations. The scaling of the QRE with system size reveals quantum phase transitions involving hidden and conventional order.
We connect QRE scaling in the weakly interacting superfluid phase to Luttinger-liquid correlations and identify distinct critical signatures in its extensive and subleading contributions. 
At the Mott–Haldane transition, subtracting the extensive background reveals a pronounced peak in the subleading contribution, providing a clear signature of the onset of hidden string order.
At the Haldane–density-wave transition, the extensive contribution to the QRE instead exhibits scaling consistent with Ising criticality.
These results establish non-Gaussianity as a probe of many-body criticality, linking bosonic quantum resources to strongly correlated matter.
\end{abstract}
\maketitle

\paragraph{Introduction.}
Quantum resource theories~\cite{preskill2012quantum, chitambar2019,daley2022practical} provide a framework for quantifying the properties of quantum states relative to a specified set of control and measurement operations. The choice of free states and operations determines which processes are accessible and which require additional resources.
Entanglement theory is a central example: separable states are free under local operations and classical communication, while entangled states enable otherwise inaccessible tasks ~\cite{horodecki2009,vidal2003}. Resource theories also illuminate the limits of classical simulation. Limited entanglement can allow efficient tensor-network representations ~\cite{verstraete2004,orus2019tensor,ran2020tensor,banuls2023tensor}, but entanglement alone does not guarantee computational advantage of quantum systems. Highly entangled stabilizer states remain efficiently simulable under Clifford operations~\cite{gottesman1998heisenberg,aaronson2004improved}. This distinction motivates the study of non-stabilizerness, or “magic,” as a resource beyond entanglement~\cite{veitch2012negative,howard2017,wang2019quantifying,liu2022,leone2022}.

For bosonic systems, Gaussian states provide an important reference class. They are fully characterized by the mean values and covariance matrix of the canonical quadrature operators, allowing a compact description even for large systems~\cite{Weedbrook2012}. This structure underlies the correlation-matrix treatment of quadratic bosonic models and enables efficient classical simulation of processes involving Gaussian states, operations, and measurements~\cite{bartlett2002, Matsuda26unifying}. Going beyond these restrictions motivates the study of non-Gaussian states and operations as resources for quantum information processing~\cite{Takagi2018,Albarelli2018}.

Beyond the computational role, quantum resources offer a perspective on collective behavior in many-body systems. Entanglement has become a central diagnostic of ground-state phases, quantum criticality~\cite{Amico2008,gogolin2016equilibration,eisert2010,calabrese2004entanglement,laflorencie2016quantum}, and nonequilibrium dynamics~\cite{eisert2015quantum,abanin2019, Sierant25mbl}. More recently, studies of non-stabilizerness have explored how resources beyond entanglement are organized across phases~\cite{white2021,haug2023,tarabunga2023,odavic2023complexity,santra2025complexity, Hoshino26cftSRE} and evolve under quantum dynamics~\cite{Turkeshi25spread, rottacaso2023,Turkeshi2025,Tirrito2025,Odavic2025,haug2025probing,passarelli2025,loio2026,falcao2025,Bejan2026,tirrito2025universal}. 
A complementary perspective has emerged in fermionic systems~\cite{Sierant2026, Tarabunga2026, Hoshino26fermio,coffman2025measuring,Lumia2024, Lyu24, Hebenstreit2019,Ares2026asymmetry,Ares2026random}, where non-Gaussianity has been used to probe ground-state quantum criticality and  quantum quenches~\cite{Aditya2025,falcao2026fermionic,bhakuni2026quantum}.
These developments raise a corresponding question for interacting bosons: 

\textit{How does non-Gaussianity encode long-distance correlations and quantum phase transitions?}

\begin{figure}[t]
 \centering
 \includegraphics[width=\columnwidth]{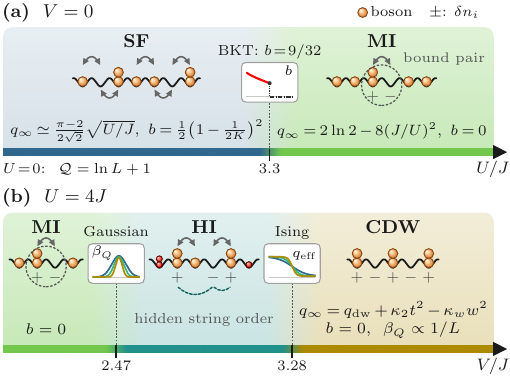}
 \caption{\textbf{Structure of the QRE across the phases of the extended Bose--Hubbard chain.} For a chain of length $L$, the QRE contains an extensive contribution $Lq_\infty$ and a possible logarithmic correction $b\ln L$. The finite-size estimators $\beta_Q$ and $q_{\rm eff}$ probe the subleading and extensive contributions, respectively. (a) At $V=0$, the superfluid exhibits a logarithmic contribution that vanishes in the Mott insulator beyond the BKT transition ($U_c\simeq3.3J$). The value $b=9/32$ is obtained by evaluating the weak-coupling reference at $K=2$. (b) At $U=4J$, the Mott--Haldane ($V\simeq2.47J$) and Haldane--density-wave ($V\simeq3.28J$) transitions manifest via a peak of $\beta_Q$ and a crossing of $q_{\rm eff}$. Cartoons: bosons in lattice wells, hopping (grey arrows), and $\delta n_i=n_i-1$ ($\pm$). 
 }
 \label{fig:sketch}
\end{figure}
We address this question using the quantum relative entropy (QRE)~\cite{Genoni2008} between a many-body state and its closest Gaussian reference, a measure that for pure states can be efficiently computed and experimentally inferred from first and second moments of the canonical quadratures.
Combining analytical calculations and density-matrix renormalization group (DMRG) simulations~\cite{White1992, Schollwock2011, itensor}, we study the QRE in the extended Bose–Hubbard chain at unit filling~\cite{Lahaye09, Goral02, Dutta15review, Biedron18ext, Lacki14, Lacki2016}. 
We connect weak-coupling behavior of QRE to Luttinger-liquid correlations, and track the subleading terms across the superfluid–Mott transition. 
At the Mott–Haldane transition, the subleading term develops a peak that locates the transition more robustly than string-order crossings, while the extensive response at the 
Haldane–density-wave transition follows scaling 
consistent with Ising criticality.  
Our results show that bosonic magic resources, quantified by the QRE, reflect collective behavior and quantum criticality in interacting many-body systems.
Fig.~\ref{fig:sketch} provides a schematic overview of this physical picture.

\paragraph{The model.} We consider the one-dimensional extended Bose--Hubbard model on an open chain of $L$ sites~\cite{Berg2008,Rossini2012},
\begin{equation}
 \begin{aligned}
 H={}&-J\sum_{i=1}^{L-1}(b_i^\dagger b_{i+1}+\mathrm{H.c.})\\
 &+\frac{U}{2}\sum_{i=1}^{L}n_i(n_i-1)
   +V\sum_{i=1}^{L-1}n_i n_{i+1},
 \end{aligned}
 \label{eq:hamiltonian}
\end{equation}
where $b_i$ annihilates a boson on site $i$, $n_i=b_i^{\dagger}b_i$, $J$ is the hopping amplitude, and $U$ and $V$ are the on-site and nearest-neighbour interaction strengths. The Hamiltonian conserves the particle number $N=\sum_i n_i$ and is invariant under lattice inversion. We set $J=1$ and work at unit filling, $N=L$.
At weak coupling, the superfluid (SF) is a gapless Luttinger liquid with correlations $C(r)=\langle b_i^\dagger b_{i+r}\rangle\propto r^{-1/(2K)}$. For $V=0$, increasing $U$ drives a Berezinskii–Kosterlitz–Thouless transition into the Mott insulator (MI) at $U_c/J\simeq3.3$, where the Luttinger $K$ parameter reaches
the universal value $K_c=2$~\cite{kuhner1998,Kuhner2000, Thamm25bkt, Rams18}. At finite $V$, the system also supports Haldane (HI) and charge-density-wave (CDW) insulator phases~\cite{DallaTorre2006,Berg2008,Rossini2012}. In the MI, occupation fluctuations form bound particle–hole pairs around a singly occupied background. The HI exhibits hidden string order: excess particles and holes alternate after disregarding singly occupied sites, without local symmetry breaking. The CDW instead exhibits alternating high and low site occupations, spontaneously breaking translation symmetry by one lattice site in the bulk. Together, these phases provide a setting for studying non-Gaussianity across transitions involving hidden string order and conventional density order.

\paragraph{Non-Gaussianity in bosonic systems.}
Any bosonic state $\rho$ with finite second moments defines a unique Gaussian reference $\rho_G$: the Gaussian state with the same first moments and covariance matrix. 
This reference maximizes the entropy among all states sharing these moments~\cite{Wolf2006,Marian2013}.
We quantify the departure from Gaussianity by minimizing the quantum relative entropy (QRE) over the set $\mathcal G$ of Gaussian states. The minimum is attained by the reference $\rho_G$ with matching moments, giving
\begin{equation}
 \Q(\rho)=S(\rho\Vert\rho_G)=S(\rho_G)-S(\rho),
 \label{eq:qre_definition}
\end{equation}
where $S(\rho)=-\Tr(\rho\ln\rho)$ and $S(\rho\Vert\sigma)=\Tr[\rho(\ln\rho-\ln\sigma)]$. The entropy difference follows because $\rho$ and $\rho_G$ share the same first and second moments~\cite{Genoni2008,Genoni2010}. The QRE $\Q(\rho)$ vanishes for a Gaussian state and is positive otherwise.
It is additive on tensor-product states, invariant under Gaussian unitaries, and nonincreasing under Gaussian quantum channels~\cite{Genoni2008,Genoni2010}.

For states with fixed total particle number, $\langle b_i\rangle=\langle b_i b_j\rangle=\langle b_i^\dagger\rangle=\langle b_i^\dagger b_j^\dagger\rangle=0$. The Gaussian reference is therefore completely determined by the one-body correlation matrix $C_{ij}=\langle b_i^\dagger b_j\rangle $. 
Its diagonal elements are site occupations and its off-diagonal elements describe coherence between sites. The eigenvalues $n_\alpha$ of $C_{ij}$ are the natural occupations, with $\sum_\alpha n_\alpha=N$. For a pure many-body state, $S(\rho)=0$, so its non-Gaussianity is
\begin{equation}
    \Q(\psi)=\sum_\alpha g(n_\alpha), \quad g(n)=(n+1)\ln(n+1)-n\ln n
    \label{eq:gaussian_entropy}
\end{equation}
with $g(0)=0$ and natural logarithms throughout. Here $g(n)$ is the entropy of a thermal mode with mean occupation $n$. 
For pure states with fixed particle number, the natural occupations $n_\alpha$ determine $\mathcal Q$ completely.  
At unit filling, $\sum_\alpha n_\alpha=L$, concavity of the single-mode entropy in Eq.~\eqref{eq:gaussian_entropy} gives the exact bounds
\begin{equation}
(L+1)\ln(L+1)-L\ln L \leq \Q \leq 2L\ln2.
\label{eq:qre_bounds}
\end{equation}
The lower bound is attained when all particles occupy one orbital. The product Mott state $\lvert1\rangle^{\otimes L}$ attains the upper bound despite having no intersite entanglement. 
The Mott-state example shows that a large QRE can arise from local number quantization alone. Its magnitude is therefore insufficient to identify collective critical behavior. We instead study how \(\mathcal Q\) changes with system size, separating its leading extensive term from the finite-size corrections:
\begin{equation}
 \Q(L)=Lq_\infty+b\ln L+c_0+\frac{a}{L}+\cdots,
 \label{eq:qre_expansion}
\end{equation}
where $q_\infty$ is the QRE density, $b$ describes a possible logarithmic contribution, and $c_0$ and $a/L$ account for constant and inverse-size corrections. 
Probing the dependence of the coefficients $q_\infty, b$ on the interaction strengths $U$ and $V$ allows us to identify which scaling contributions carry signatures of the phase transitions.
In particular, to isolate the subleading response, we combine QRE values for lengths $(L/4,L/2,L)$,
\begin{equation}
 \beta_Q(L)=
 \left[3\Q(L/2)-2\Q(L/4)-\Q(L)\right]/\ln2.
 \label{eq:beta_q}
\end{equation}
This combination cancels both the extensive term and the size-independent offset in Eq.~\eqref{eq:qre_expansion}, yielding $\beta_Q(L)=b-3a/(L\ln2)+\cdots$. Thus, $\beta_Q$ approaches the logarithmic coefficient $b$ in the large-system limit.
To extract the extensive contribution, we introduce the complementary finite-size estimator:
\begin{equation}
 q_{\mathrm{eff}}(L)=
 4\left[\Q(L)-2\Q(L/2)+\Q(L/4)\right]/L.
 \label{eq:qre_leading}
\end{equation}
This combination cancels the logarithmic and constant terms in Eq.~\eqref{eq:qre_expansion}, yielding $q_{\mathrm{eff}}=q_\infty+4a/L^2+\cdots$. Together, $\beta_Q$ and $q_{\mathrm{eff}}$ allow us to examine the subleading and extensive responses across the phase transitions.

\begin{figure}[t]
 \centering
 \includegraphics[width=0.45\textwidth]{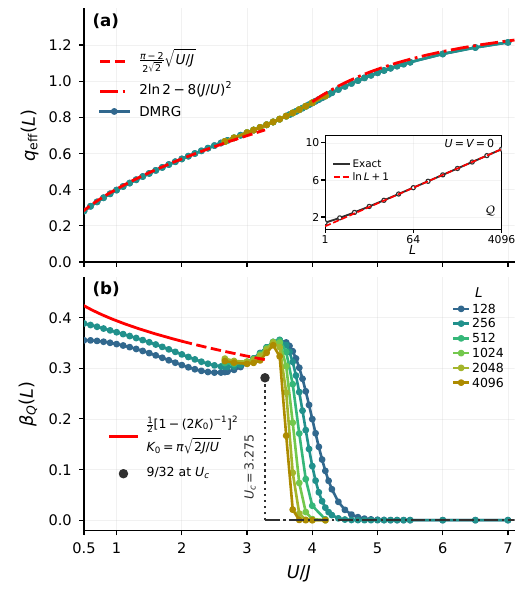}
\caption{\textbf{Non-Gaussianity across the superfluid--Mott transition.}
Results for the one-dimensional Bose--Hubbard model at $V=0$, with open boundaries and unit filling $N=L$. (a) Extensive QRE estimator $q_{\mathrm{eff}}(L)$. Red curves show the weak-coupling expression and the strong-coupling expansion. In the inset, the exact QRE at $U=V=0$ is compared with its large-$L$ asymptote $\ln L+1$.
(b) Subleading estimator $\beta_Q(L)$. The red curve shows the weak-coupling prediction$ \frac12[1-(2K_0)^{-1}]^2$, with $K_0=\pi\sqrt{2J/U}$. The black point marks the reference value $9/32$ obtained by setting $K=2$ at $U_c/J=3.275$. Black dotted guides indicate the SF-MI transition. 
The finite-size results show a rounded crossover toward a vanishing subleading response, while $q_{\mathrm{eff}}$ remains smooth.}\label{fig:qre_sf_mi}
\end{figure}
\paragraph{Weakly interacting limit.}
For noninteracting bosons, $U=V=0$, all $N$ particles occupy the lowest single-particle orbital of the Hamiltonian \eqref{eq:hamiltonian}.
The ground state is therefore the Fock state $\lvert N\rangle$ of this orbital, with all other orbitals empty. Its Gaussian reference is thermal state in the occupied orbital, with mean occupation $N$.
At unit filling, Eq.~\eqref{eq:gaussian_entropy} gives $\mathcal Q=g(N)=\ln L+1+\frac{1}{2L}+\cdots$, cf. the inset in Fig.~\ref{fig:qre_sf_mi}(a).
Thus, the coefficients in Eq.~\eqref{eq:qre_expansion} are $q_\infty=0$, $b=1$, and $c_0=1$. This non-Gaussianity of the $\lvert N\rangle$ state originates entirely from the sharp particle number constraint and is independent of the spatial profile of the occupied orbital.

Within the SF phase, weak interactions populate additional natural orbitals and generate an extensive contribution to the QRE. 
Number-conserving Bogoliubov theory~\cite{Castin1998} gives the condensate occupation $N_0$ and the occupations $v_k^2$ of noncondensed momentum modes, yielding $\mathcal Q=g(N_0)+\sum_{k\neq0}g(v_k^2)$ (see  \hyperref[app:weak_int]{Appendix}). 
Below the inverse healing length, $\xi_h=\sqrt{J/[2(U+2V)]}$, the depleted occupations behave as $v_k^2\simeq1/(4|k|\xi_h)$. Their contribution gives the leading weak-coupling QRE density,
\begin{equation}
  q_\infty \simeq \frac{\pi-2}{2\sqrt{2}} \sqrt{\frac{U+2V}{J}}.
  \label{eq:qsf}
\end{equation} 
Fig.~\ref{fig:qre_sf_mi} (a) compares this analytical prediction with DMRG results at $V=0$. The estimators $q_{\mathrm{eff}}(L)$ and $\beta_{\mathcal Q}(L)$, probe the leading extensive and subleading contributions to the QRE. The extensive estimator $q_{\mathrm{eff}}(L)$ follows the square-root dependence at weak coupling and approaches $q_\infty=2\ln2-8(J/U)^2$ as $U/J$ increases, with little size dependence. The extensive response remains smooth across the SF–MI transition. 

The subleading logarithmic term reflects competing contributions from the largest natural occupation and the remaining modes. Luttinger-liquid correlations give  the condensate population $N_0\propto L^{1-1/(2K)}$, with $K=\pi\sqrt{2J/(U+2V)}$ at weak coupling. Since $g(n)\simeq\ln n+1$ for large $n$, this contributes $[1-1/(2K)]\ln L$. 
For open chains at weak coupling, the positive logarithmic contribution from the largest natural occupation outweighs the negative correction from the remaining modes, leaving a positive net coefficient.
Fig.~\ref{fig:qre_sf_mi} (b) shows the corresponding DMRG estimates of $\beta_{\mathcal Q}(L)$. At weak coupling, increasing chain length brings the results toward the reference $\tfrac12[1-(2K_0)^{-1}]^2$, where $K_0=\pi\sqrt{2J/U}$, consistent with $b\simeq\tfrac12-\tfrac{1}{2K_0}$ to leading order in $1/K_0$. Near the SF–MI transition, a nonmonotonic crossover of $\beta_{\mathcal Q}(L)$ precedes a rapid decrease toward zero in the insulating regime. This decrease sharpens with size but retains appreciable finite-size drift.

\begin{figure*}[t]
\centering
\includegraphics[width=\textwidth]{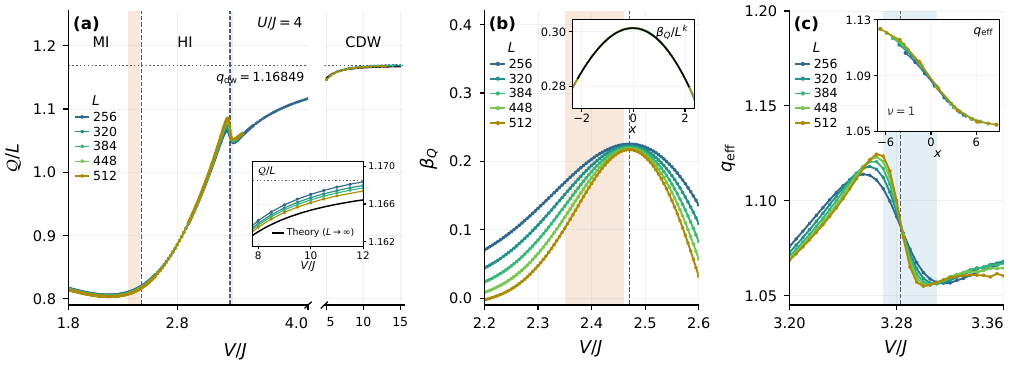}
\caption{ \textbf{Non-Gaussianity across the MI--HI and HI--CDW phase transitions at $U/J=4$, unit filling, and open boundaries.} Colors consistently identify $L=256,320,384,448,512$. (a) QRE density $\mathcal{Q}/L$. The $L=256$ curve extends to $V/J=4$; the larger sizes stop at $V/J=3.4$. Beyond the axis break, the black curve shows the analytical thermodynamic density from Eq.~\eqref{eq:lv_free} for $5\leq V/J\leq15$, approaching the dotted asymptote
$q_{\mathrm{dw}}\simeq1.168$.
The two horizontal-axis segments use different scales. (b) Subleading response $\beta_Q$ near MI--HI. Inset: collapse under $x=(V/J-2.471)L^{0.66}$ and $\beta_Q/L^k$, with $k=-0.0525$; the black line is the mean of the rescaled curves. (c) Extensive coefficient $q_{\mathrm{eff}}$ near HI--CDW. Inset: direct rescaling of the results for
$3.27\leq V/J\leq3.30$ against
$x=(V/J-3.283)L$, fixing the Ising exponent $\nu=1$. Dashed vertical lines mark the QRE transition estimates. The orange band shows the string-order extrapolation range $2.35\leq V/J\leq2.46$ discussed in the Appendix; the blue band shows the HI--CDW reference interval
$3.27\leq V/J\leq3.31$.
}
\label{fig:qre_transitions}
\end{figure*}

\paragraph{Insulating transitions.}
We now fix $U/J=4$ and increase the nearest-neighbour repulsion $V$, traversing the MI, HI, and CDW phases. Fig.~\ref{fig:qre_transitions} shows the QRE density obtained from DMRG. It has a broad minimum within the MI and increases smoothly through the MI–HI transition. Near the HI–CDW transition, a maximum is followed by a decrease and renewed growth in the CDW phase. 
To understand non-Gaussianity across these phases and their transitions, we analyze its extensive and subleading contributions separately.

At the Mott--Haldane transition, we examine the subleading response through the estimator $\beta_Q$. While the QRE density varies smoothly [Fig.~\ref{fig:qre_transitions}(a)], $\beta_Q$ develops a pronounced peak near the transition [Fig.~\ref{fig:qre_transitions}(b)]. The peak height changes only weakly with system size, whereas its narrowing provides 
a clear signature of criticality. The peak profiles collapse under
\begin{equation}
 \beta_Q(L,V)=L^k\Phi\!\left[(V/J-V_Q/J)L^y\right].
 \label{eq:mi_hi_qre_scaling}
\end{equation}
with $V_{\mathcal Q}/J\simeq2.471$, $y\simeq0.66$, and $k\simeq-0.0525$, for $L\in[256,512]$. The inset of Fig.~\ref{fig:qre_transitions} (b) shows the common rescaled profile. The peak position exhibits substantially weaker finite-size drift and less sensitivity to extrapolation than transition estimates extracted from string-order correlations~\cite{DallaTorre2006,Berg2008} evaluated on the same ground states (see \hyperref[app:stability]{Appendix}).
The narrowing of the peak connects this response to the critical correlations. Since $\xi\propto|V/J-V_c/J|^{-\nu}$, finite-size rounding occurs over a window proportional to $L^{-1/\nu}$, motivating $y=1/\nu$. At the MI–HI transition, Luttinger theory gives $\nu^{-1}=2-K$ and the string-correlation exponent $\eta_s=K/2$ ~\cite{Berg2008,Cazalilla2011}. 
The numerical estimate $\eta_s\simeq0.70$ implies \(K\simeq1.4\) and \(1/\nu\simeq0.60\), close to the QRE collapse exponent. 
The largest natural occupation, i.e., the largest eigenvalue of the one-body correlation matrix $C$, provides a consistent estimate: its growth as approximately \(L^{0.64}\) between \(L=256\) and \(512\), together with \(n_{\mathrm{largest}}\propto L^{1-1/(2K)}\), again gives \(K\simeq1.4\). This value also agrees with previous results at \(U/J=4\)~\cite{Deng2013}.

Increasing nearest-neighbour interaction strength $V$ further leads to the HI--CDW transition, where staggered density order emerges through twofold symmetry breaking. This transition belongs to the Ising universality class. Here, the extensive estimator $q_{\mathrm{eff}}$, defined in Eq.~\eqref{eq:qre_leading}, provides a clear transition signature: curves for different system sizes cross within a narrow coupling region [Fig.~\ref{fig:qre_transitions}(c)]. The crossing between the two largest sizes, $L=448$ and $512$, yields the finite-size transition estimate $V_\times/J\simeq3.28312$. Using this crossing as the center, we test Ising scaling by plotting the results against \(x=(V/J-V_\times/J)L\). The curves closely align over \(3.27\leq V/J\leq3.30\), without rescaling their vertical values or subtracting a background (inset of Fig. 3(c)). The extensive response therefore exhibits finite-size scaling consistent with Ising criticality, complementing the subleading signature of the MI–HI transition.

\paragraph{Strongly interacting limit.}
Beyond the HI-CDW transition, the QRE density increases toward a strong-coupling limit (see the inset of Fig.~\ref{fig:qre_transitions} (a)).  Deep in the CDW phase the coefficients of Eq.~\eqref{eq:qre_expansion} can be obtained with a strong coupling limit $J\ll V-U$ calculation. 
In this limit, the ground state features a delocalized domain wall separating the two CDW orientations (see \hyperref[app:lv]{Appendix}).
A perturbative treatment of its motion and the virtual hopping processes gives:
\begin{equation}
  q_\infty=q_{\rm dw}+\kappa_2t^2-\kappa_w w^2+\cdots,\;\;\;
  q_{\rm dw}=\int_0^1\!ds\,g\big(1+D(s)\big),
  \label{eq:lv_free}
\end{equation}
where $D(s)=1-2s+\sin(2\pi s)/\pi$ is the staggered density profile and $s\in[0,1]$ is the normalized position along the chain., $t=\frac{2J}{3V-U}$, and $w=\frac{6J^2(6V-U)}{U(3V-U)^2}$.
The coefficients and derivation are discussed in the \hyperref[app:lv]{Appendix}. The $t^2$ term gives the leading hopping correction, while the $w^2$ term incorporates a selected higher-order contribution from the broadening of the natural-occupation spectrum. 
The inset of Fig.~\ref{fig:qre_transitions}(a) shows the QRE densities approaching this thermodynamic prediction as $L$ increases.

\paragraph{Conclusion and outlook.}
In this work, we characterized bosonic magic resources in the extended Bose–Hubbard model through the lens of the QRE. 
Our findings connect the extensive and subleading contributions of QRE to Luttinger-liquid correlations, quantum criticality, and strong-coupling domain-wall physics. The subleading response resolves the superfluid–Mott crossover and develops a peak at the Mott–Haldane transition with substantially weaker finite-size drift and extrapolation dependence than string-order crossings. At the Haldane–density-wave transition, the extensive response exhibits scaling consistent with Ising criticality. 
For pure states with fixed particle number, these properties of non-Gaussianity are determined entirely by one-body correlations, opening a route to experimental studies of QRE in optical lattices~\cite{Gluza21,Impertro24} or in photonic systems~\cite{Eichler11}.

Our results provide a reference point for several directions of investigation of non-Gaussianity in bosonic systems. The Haldane phase findings motivate extensions to symmetry-protected topological matter~\cite{Ejima2014,Pollmann2010,Pollmann2012}. Following the QRE peak upon breaking inversion symmetry, which allows the MI and HI to connect without a sharp transition~\cite{Berg2008}, would test its sensitivity to symmetry protection. In higher dimensions, comparison with topological entanglement entropy could establish whether non-Gaussianity derived from one-body correlations retains signatures of intrinsic topological order~\cite{KitaevPreskill2006,LevinWen2006}.
Beyond equilibrium, tracking non-Gaussianity after interaction quenches could clarify its relation to relaxation, memory of initial conditions, and the spreading of correlations and entanglement~\cite{Kollath07,Lauchli08}. Its growth and saturation could also be investigated in regimes exhibiting long-lived metastability~\cite{Carleo12}, many-body localization~\cite{Sierant17, Sierant18, Orell19}, and slow dynamics in tilted lattices~\cite{Yao2020}. Lattice gauge theories offer a complementary direction~\cite{Banuls2020,Zohar2016}: Bose–Hubbard gauge simulators provide a setting for comparing QRE with gauge-invariant observables during flux-string formation and breaking~\cite{Yang2020}. Developing a reference that respects Gauss-law constraints could help separate their contribution from interaction-generated correlations.
Finally, dipolar lattice bosons offer a natural extension of the present analysis, since their interactions reach beyond nearest neighbours and stabilize additional ordered phases~\cite{Goral02,Lacki24ground}. In optical ladders, the long-range tail produces a topological floating phase, a critical phase with incommensurate density correlations that separates a crystalline phase from the Haldane phase~\cite{Korbmacher25floating}. 
Following the QRE across this gapless intermediate regime would test how non-Gaussianity responds to longer-range interactions and to topology without a bulk gap, and would connect bosonic magic resources to platforms already under experimental investigation.

\begin{acknowledgments}
\paragraph{Acknowledgments}
We acknowledge insightful discussions with Yuto Ashida.
P. Stornati acknowledges funding from the project PID2025-172563NA-I00, funded by MICIU/AEI/10.13039/501100011033 and by ERDF/EU. P. Sierant acknowledges fellowship within the “Generación D” initiative, Red.es, Ministerio para la Transformación Digital y de la Función Pública, for talent attraction (C005/24-ED CV1), funded by the European Union NextGenerationEU funds, through PRTR.
E.T. was funded by the Swiss National Science Foundation (SNSF) under Grant No.  MPFP2\_234754. E.T. acknowledges CINECA (Consorzio Interuniversitario per il Calcolo Automatico) award, under the ISCRA initiative and Leonardo early access program, for the availability of high-performance computing resources and support.

{\it AI Statement.--} Claude (Anthropic) and ChatGPT (OpenAI) have been used to check the derivations, to  assist with development of codes, and to improve the writing. Any AI-generated content has been verified by the authors, who take full responsibility for it.
\end{acknowledgments}

\paragraph{Data availability}
The code and the data for our simulations will be publicly shared at publication.

\label{page:main-end}
\onecolumngrid
\vspace{0.5cm}

\twocolumngrid
\phantomsection\label{page:references-start}
\bibliography{references}
\onecolumngrid
\clearpage
\twocolumngrid

\raggedbottom

\appendix
\setcounter{figure}{0}
\renewcommand{\thefigure}{A\arabic{figure}}
\renewcommand{\theHfigure}{appendix.\arabic{figure}}
\setcounter{equation}{0}
\renewcommand{\theequation}{A\arabic{equation}}
\renewcommand{\theHequation}{appendix.\arabic{equation}}

\section*{QRE in the weakly interacting superfluid}
\label{app:weak_int}
\subsection{Bogoliubov occupations}
We treat Eq.~\eqref{eq:hamiltonian} at unit filling, $N=L$, in number-conserving Bogoliubov theory~\cite{Castin1998}.  
On a ring the ground state is invariant under translations, so the one-body correlation matrix $C_{ij}=\langle c_i^\dagger c_j\rangle$ depends only on $j-i$ and its eigenvectors are plane waves.  The natural occupations are therefore the momentum occupations $n_k=\langle a_k^\dagger a_k\rangle$, with $a_k=L^{-1/2}\sum_j e^{-ikj}c_j$ and $k=2\pi m/L$.  The largest is the condensate, $N_0=n_{k=0}$, and Bogoliubov theory gives $n_k=v_k^2$ for $k\neq0$, with
\begin{equation}
  v_k^2=\frac12\Big[\frac{\varepsilon_k+\tilde V_k}{E_k}-1\Big],\quad
  E_k=\sqrt{\varepsilon_k(\varepsilon_k+2\tilde V_k)},
  \label{eq:vk}
\end{equation}
$\varepsilon_k=2J(1-\cos k)$, $\tilde V_k=U+2V\cos k$, and $N_0=N-\sum_{k\neq0}v_k^2$.  Since the state is pure and number conserving, Eq.~\eqref{eq:gaussian_entropy} gives, at this level,
\begin{equation}
  \mathcal{Q}_{\rm Bog}=g(N_0)+\sum_{k\neq0}g(v_k^2).
  \label{eq:qbog}
\end{equation}
For $k\ll1/\xi_h$, $v_k^2=\kappa/|k|-\tfrac12+O(k)$ with $\kappa=\sqrt{(U+2V)/8J}=1/4\xi_h$.  We write $\eta\equiv\kappa/\pi=1/2K$, with $K=\pi\sqrt{2J/(U+2V)}$ the weak-coupling
Luttinger parameter.

\subsection{Thermodynamic density}
With $x=k\xi_h$ and $\varepsilon_k\simeq Jk^2$, Eq.~\eqref{eq:vk} takes the scaling form $v^2(x)=(2x^2+1)/(4|x|\sqrt{x^2+1})-\tfrac12$, which behaves as $1/4|x|$ for $x\ll1$ and $1/16x^4$ for $x\gg1$.  Hence the leading behaviour
\begin{equation}
  q_\infty\simeq q_{\rm Bog}\equiv\int_{-\pi}^{\pi}\frac{dk}{2\pi}\,g(v_k^2)
  \simeq\frac{I}{2\pi}\sqrt{\frac{2(U+2V)}{J}},
\end{equation}
with $I=\int_{-\infty}^{\infty}dx\,g(v^2(x))=1.79321$.
We note that there are two additional corrections.  The lattice integral $q_{\rm Bog}$ exceeds the scaling form by $1\%$ at $U+2V=0.1J$ and $5\%$ at $U+2V=J$ for $V=0$, and it depends on $U$ and $V$ separately.  More importantly, $q_{\rm Bog}$ is not the exact density beyond leading order: the occupations of modes with $k\xi_h\sim1$ have relative corrections of order $1/K$.  

\subsection{Bogoliubov sum on a ring}
We write $g(v_k^2)=-\ln|2\sin(k/2)|+s(k)$.  In terms of $m_k=v_k^2+\frac12$,
$g=\ln m_k+1+O(m_k^{-2})$ contains only even powers of $1/m_k$, and
$1/m_k$ equals $|\sin(k/2)|$ times an analytic function, so $s$ is smooth,
periodic and even, with $s(0)=\ln\kappa+1$.  Since
$\prod_{m=1}^{L-1}2\sin(\pi m/L)=L$ and $\int\ln|2\sin(k/2)|\,dk=0$,
\begin{equation}
  \sum_{k\neq0}g(v_k^2)=Lq_{\rm Bog}-\ln L-\ln\kappa-1,
  \label{eq:depleted}
\end{equation}
up to exponentially small terms.  With
$\rho_{\rm reg}=\int\frac{dk}{2\pi}[v_k^2-\kappa/|k|]
\simeq-\eta[\ln(\pi\xi_h/2)+2]$,
\begin{equation}
  \frac{N_{\rm dep}}{L}=\eta\Big[\ln\frac L2+\gamma_E\Big]
  +\rho_{\rm reg}+\frac1{2L},
  \label{eq:depletion}
\end{equation}
so that
\begin{equation}
  \mathcal{Q}_{\rm Bog}=Lq_{\rm Bog}+\ln\frac{N_0}{\kappa L}+\frac{1}{2N_0}
  +O(L^{-2}),
  \label{eq:ringclosed}
\end{equation}
which direct evaluation of Eq.~\eqref{eq:qbog} confirms.
Equation~\eqref{eq:ringclosed} has an apparent logarithmic coefficient
$-\eta$.  This is not a property of the exact state, as shown below.

\subsection{Open chain}
For open boundary conditions, translational invariance is lost, and the natural occupations are obtained by diagonalizing the one-body correlation matrix constructed from the Bogoliubov--de Gennes mode functions~\cite{Dziarmaga06}. These occupations determine the QRE through Eq.~\eqref{eq:gaussian_entropy}.
The condensate is described by the lattice Gross--Pitaevskii ground state $\phi_i$, whose density is approximately uniform in the bulk and decreases smoothly to zero near each boundary as $\tanh^2(x/2\xi_h)$.
The depleted modes are the Bogoliubov-de~Gennes modes $\nu=1,\dots,L-1$, obtained from
$A=-JT+\mathrm{diag}[2U\phi_i^2+V(\phi_{i-1}^2+\phi_{i+1}^2)-\mu] +V\,T\!\circ\!\phi\phi^{T}$ and
$B=\mathrm{diag}[U\phi_i^2]+V\,T\!\circ\!\phi\phi^{T}$, with $T$ the hopping adjacency matrix and $\circ$ denoting the elementwise multiplication. In the bulk, the modes are standing waves, $k_\nu\simeq\pi\nu/L$, forming a single tower with spacing $\pi/L$, and the singularity of $g(v_k^2)$ is sampled once:
\begin{equation}
  \sum_{\nu=1}^{L-1}(-\ln k_\nu)=L(1-\ln\pi)-\tfrac12\ln L+\tfrac12\ln\tfrac\pi2.
\end{equation}
With the depletion coefficient $\eta$, this gives the coefficient $\frac12-\eta$ at the Bogoliubov level.

Accounting for Luttinger-liquid correlations, the largest natural occupation scales as $N_0\propto L^{1-\eta}$. For the remaining low-lying modes, we use the scaling form $n_\nu\propto L^{1-\eta}(\nu+\delta)^{-(1-\eta)}$, where $\delta=O(\eta)$ accounts for the boundary-induced shift of the mode indices. Summing their contributions to the QRE gives
\begin{equation}
  b_{\rm open}=(1-\eta)\Big(\frac12-\delta\Big)=\frac12-\frac{1}{2K}+O(K^{-2}).
  \label{eq:opencoeff}
\end{equation}
The last equality uses $\delta\simeq\eta/2$, which we obtain numerically and have not derived: resumming the BdG correlations as
$C_{ij}=\sqrt{n_in_j}\,e^{-D_{ij}}$, whose first order is BdG, for $L$ up to $4096$, yields $b=0.450$, $0.423$, $0.393$ and $0.353$ at $U/J=0.2$, $0.5$, $1$ and $2$, close to $(1-\eta)^2/2$.  At order $\eta$, the Bogoliubov coefficient
is thus recovered because the corrections from the exponent and from the shift
cancel.  
The constant and the $1/L$ term have no controlled weak-coupling form and are obtained from fits to numerical results.  For orientation, at fixed $N=L$ the healing layers raise the
bulk density to $n_b\simeq1+4\xi_h/L$, adding $+I/\pi$ to $\mathcal{Q}$, while a local-density estimate with density $q_\infty\sqrt{n(x)}$ removes
$2\ln2\,I/2\pi$ per edge.  The net edge term is then $-0.22$, against
$-0.72$ to $-0.60$ from the Bogoliubov-de~Gennes fits at $U/J=0.02$--$0.2$; the local-density estimate therefore does not capture the boundary constant.

\section*{QRE for strong interactions}
\label{app:lv}

\begin{figure}[t]
 \centering
 \includegraphics[width=\columnwidth]{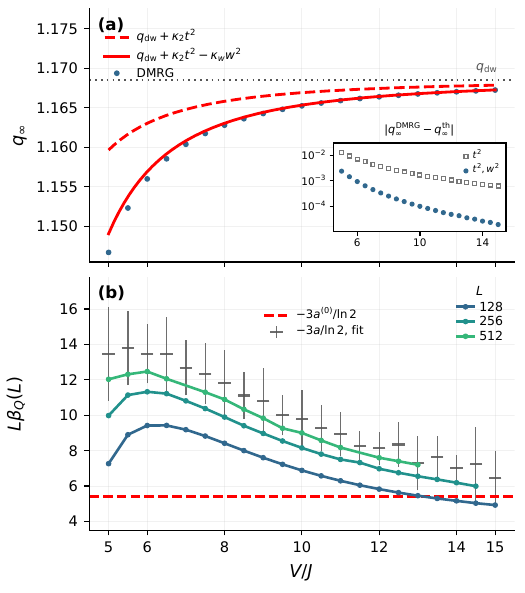}
 \caption{QRE in the CDW phase at large $V$, for $U=4J$ and open chains at unit filling. (a) Thermodynamic density $q_\infty$ extracted from DMRG fits to  $\Q=Lq_\infty+c_0+a/L+e/L^2$ with $L\ge64$ (circles). The solid and dashed curves show Eq.~\eqref{eq:lv_free} and its leading
 correction $q_{\rm dw}+\kappa_2t^2$, respectively; the dotted line marks  $q_{\rm dw}$. Inset: absolute deviations from the two analytical predictions.  (b) Rescaled subleading estimator $L\beta_Q(L)$ for the indicated chain lengths, approaching $-3a/\ln2$ as $L$ increases. Grey markers show $-3a/\ln2$ extracted from the same fits.  The dashed line marks the zeroth-order domain-wall prediction $-3a^{(0)}/\ln2$.}
 \label{fig:lv_scan}
\end{figure}

Here, our aim is to calculate $\Q(L,V)$, its thermodynamic density
$q_\infty$, and the finite-size estimator $\beta_Q$ in the large-$V$ CDW regime of Eq.~\eqref{eq:hamiltonian}.  We consider an open chain at unit filling, $N=L$ with even $N$, and write
\begin{align}
 H&=H_0+T,\\
 H_0&=\frac U2\sum_{i=1}^{L}n_i(n_i-1)
       +V\sum_{i=1}^{L-1}n_i n_{i+1},\\
 T&=-J\sum_{i=1}^{L-1}(b_i^\dagger b_{i+1}+\mathrm{H.c.}).
\end{align}
We restrict the derivation to $V>U$, the regime containing all points of the large-$V$ scan in Fig. ~\ref{fig:lv_scan}  ($U=4J$, $5\le V/J\le15$), and treat $T$ perturbatively.

The starting point is the ground-state manifold of $H_0$: the space spanned by all configurations with the lowest interaction energy. At $J=0$, $T$ vanishes and these are also ground states of $H$. On an even ring the two alternating CDW patterns, $|2020\cdots20\rangle$ and $|0202\cdots02\rangle$, have energy $UL/2$.
On an open chain the ground-state degeneracy is larger.  There are $L/2$ doubles and $L/2$ empty sites, but only $L/2-1$ empty sites are needed to separate the doubles.  The remaining empty site can be placed at either end or between two alternating regions.

An internal extra zero produces a pair $0\,0$, across which the alternating pattern switches from $2020\cdots$ to $0202\cdots$. This boundary between the two CDW orientations is called a domain wall, or kink.  Its possible positions, including the two endpoint configurations, define the orthonormal Fock states
\begin{equation}
 |x\rangle=|(20)^{x-1}(02)^{L/2-x+1}\rangle,
\end{equation}
where $x=1,\ldots,M,\quad M=L/2+1$. Every state has $L/2$ doubles and no adjacent occupied sites, so
\begin{equation}
 H_0|x\rangle=E_0|x\rangle,\qquad E_0=\frac{UL}{2}.
\end{equation}
The $V$ term vanishes in each configuration; moving the wall therefore costs no energy at $J=0$.  The ground manifold is the span of these $M$ states.

Turning on the hopping term $T$ couples the ground configurations $|x\rangle$ to excited Fock configurations $|e\rangle$ of $H_0$, with the same total particle number $N=L$.  For a hop from a double to an empty site, the matrix element is $-J\sqrt{2}\sqrt{1}=-\sqrt{2}J$.
The excitation cost depends on where the hop occurs.  Inside an alternating region, $2\,0\,2\to1\,1\,2$ removes one double, saving $U$, and creates bond energies $V+2V$.  A hop into the empty pair at the wall, $2\,0\,0\to1\,1\,0$, also saves $U$ but creates only one bond energy $V$.  We therefore define
\begin{equation}
 \Delta_1=3V-U,\qquad \Delta_2=V-U.
\end{equation}
These positive energy costs appear in the perturbative denominators: the first-order admixture of $|e\rangle$ is proportional to $\langle e|T|x\rangle/(E_0-E_e)$, and hence to $J/\Delta_1$ or $J/\Delta_2$ for these processes.

A single hop takes $|x\rangle$ outside the ground manifold, but two successive hops can return to it, either restoring the original configuration or moving the wall to a neighbouring position.  
At second order, these processes shift the diagonal energies and couple different wall positions.  The effective Hamiltonian derived below
then selects a coherent superposition
\begin{equation}
 |\Psi_{\rm wall}^{(0)}\rangle=\sum_x\psi_x|x\rangle,
 \qquad \sum_x|\psi_x|^2=1.
\end{equation}
Its lowest-energy amplitudes form an approximately sine-shaped standing wave, with the wall probability $|\psi_x|^2$ largest near the centre and suppressed near the ends.  Although this selection occurs through an effective Hamiltonian of order $J^2$, it fixes the leading state within the degenerate manifold.  Thus ``zeroth order''
means the limit $J\to0^+$ of the selected finite-chain ground state; at exactly $J=0$, the coefficients $\psi_x$ remain undetermined.

For the bulk correlation matrix and its QRE corrections, we will use the dimensionless combinations
\begin{equation}
 t=\frac{2J}{\Delta_1},\qquad
 w=\frac{6J^2(6V-U)}{U\Delta_1^2}.
 \label{eq:lv_tw}
\end{equation}
Here $t$ measures the leading nearest-neighbour coherence, while $w$ controls the broadening of the large natural occupations.  Their roles are derived explicitly in the bulk calculation below.

The calculation thus proceeds from the degenerate configurations of $H_0$, to the wall state selected by weak hopping, and finally to its one-body density matrix $C_{ij}=\langle b_i^\dagger b_j\rangle$. For a pure number-conserving state the QRE is
\begin{equation}
 \Q=\Tr g(C)=\sum_\alpha g(n_\alpha),
\end{equation}
where $g(n)=(n+1)\ln(n+1)-n\ln n$ and $n_\alpha$ are the eigenvalues of $C$ and $g(0)=0$.  

To organize the size dependence, we use the large-$L$ expansion of
Eq.~\eqref{eq:qre_expansion},
\begin{equation}
 \Q(L,V)=Lq_\infty(V)+b(V)\ln L+c_0(V)
 +\frac{a(V)}{L}+\cdots.
 \label{eq:lv_size_overview}
\end{equation}
Here $q_\infty(V)=\lim_{L\to\infty}\Q(L,V)/L$ is the thermodynamic QRE density, $b$ is the coefficient of a possible logarithmic term, $c_0$ is the constant term, and $a$ controls the leading inverse-size term.  This is an asymptotic expansion, rather than an exact identity at finite $L$; the leading wall calculation below gives $b=0$.
The strong-coupling calculation of the extensive coefficient starts from
\begin{equation}
 q_\infty=q_{\rm dw}+\delta q,\qquad
 \delta q\simeq\kappa_2t^2-\kappa_w w^2.
 \label{eq:lv_density_overview}
\end{equation}
The quantity $q_{\rm dw}$ is the zeroth-order density of the
hopping-selected, delocalized wall state.  The term $\kappa_2t^2$ is its first nonzero correction, of order $J^2$ at fixed $U,V$. The $-\kappa_w w^2$ term is a selected band-broadening contribution of
order $J^4$, not the complete correction at that order.  
In the following section we derive
$q_{\rm dw}$ and the hopping corrections.  Separately, the finite-chain zeroth-order profile determines $c_0^{(0)}$ and $a^{(0)}$.

\subsection{Effective Hamiltonian of the wall}
For the appropriate ground manifold defined above, let
\begin{equation}
 P=\sum_{x=1}^{M}|x\rangle\langle x|,\qquad Q=\mathbb I-P.
\end{equation}
Here $Q$ is a projector, distinct from the QRE $\Q$.  Both subspaces have $N=L$, because $T$ conserves particle number.  Writing the full eigenstate as $|\Psi\rangle=|\psi_P\rangle+|\psi_Q\rangle$, projection of the Schr\"odinger equation gives
\begin{align}
 (E-E_0)|\psi_P\rangle&=PTP|\psi_P\rangle+PTQ|\psi_Q\rangle,\\
 (E-QHQ)|\psi_Q\rangle&=QTP|\psi_P\rangle.
\end{align}
Eliminating $|\psi_Q\rangle$ and keeping two powers of $T$ gives
\begin{equation}
 H_{\rm eff}=E_0P+PTP+PTQ\frac{1}{E_0-QH_0Q}QTP.
\end{equation}
A single hop connects no two states within the kink ground manifold, so $PTP=0$.  If $H_0|e\rangle=E_e|e\rangle$ in the complementary subspace, this becomes
\begin{equation}
 \langle y|H_{\rm eff}|x\rangle=E_0\delta_{xy}
 +\sum_{e\in Q}\frac{\langle y|T|e\rangle\langle e|T|x\rangle}{E_0-E_e}.
\end{equation}
Only intermediate states reachable by one hop contribute.
Writing $\epsilon_{\rm bulk}$ for the common interior energy shift, the resulting operator is
\begin{align}
 H_{\rm eff}={}&(E_0+\epsilon_{\rm bulk})P\nonumber\\
 &-|\tau|\sum_{x=1}^{M-1}(|x+1\rangle\langle x|+\mathrm{H.c.})\nonumber\\
 &+W(|1\rangle\langle1|+|M\rangle\langle M|).
\end{align}
Its lowest eigenvector defines
\begin{equation}
 |\Psi_{\rm wall}^{(0)}\rangle=\sum_x\psi_x|x\rangle,
 \qquad \sum_x|\psi_x|^2=1.
\end{equation}
The common energy does not affect these amplitudes.
For an interior kink there are $L-4$ virtual hops of cost $\Delta_1$ and two of cost $\Delta_2$; at an endpoint there are $L-2$ and one, respectively.  Thus
$\epsilon_{\rm bulk}=-2J^2(L-4)/\Delta_1-4J^2/\Delta_2$ and
$W=2J^2/\Delta_2-4J^2/\Delta_1$.

Second-order degenerate perturbation theory in $J$ within each manifold gives a tight-binding chain for the wall position $x=1,\dots,M$, with hopping $-|\tau|$, a bulk on-site energy, and a different on-site energy at
the two ends.  For the kink, the on-site terms count the virtual single hops: every double has two hops of cost $\Delta_1$, except the two end doubles (one hop each) and the two doubles adjacent to the wall, whose hop into the wall costs $\Delta_2$.  Moving the
wall by one cell, $(0\,2)\to(2\,0)$, is a two-hop process through the intermediate `$1\,1$' of cost $\Delta_2$, with amplitude $(-\sqrt2J)^2/(-\Delta_2)=-2J^2/\Delta_2$.

\subsection{Zeroth-order QRE of the free wall}
We now use the wall ground state  to calculate the three leading size-dependent contributions,
\begin{equation}
\Q^{(0)}(L)=Lq_{\rm dw}+c_0^{(0)}+\frac{a^{(0)}}L+O(L^{-2}).    
\end{equation}
We first obtain the density profile and its thermodynamic QRE density $q_{\rm dw}$, then retain the finite-grid corrections that determine $c_0^{(0)}$ and $a^{(0)}$. Here zeroth order means the limit $J\to0^+$ of the ground state selected by the effective wall Hamiltonian, at fixed finite $L$.

The state used here is the superposition $|\Psi_{\rm wall}^{(0)}\rangle=\sum_x\psi_x|x\rangle$ found in the previous subsection.  At $J=0$, all wall positions have the same energy, so $H_0$ alone does not determine the coefficients $\psi_x$.
A small positive hopping selects them through the effective wall Hamiltonian.  We call this the zeroth-order state because we retain that selected superposition before adding the small admixtures of excited configurations produced by virtual boson hops.

For the kink, site $2j-1$ of cell $j$ is doubly
occupied when the wall lies to its right, $m\ge j$, and empty otherwise. Hence $\langle n_{2j-1}\rangle=2P_j$ and $\langle n_{2j}\rangle=2-2P_j$, with $P_j=\sum_{m=j}^{L/2}|\psi_{m+1}|^2$.  Here $m=x-1$ counts the $20$ cells to the left of the wall.  To pass to a continuum
description, introduce the rescaled wall coordinate and grid spacing
\begin{equation}
    u_m=\frac{m+x_0}{\ell},\qquad \Delta u=\frac1\ell,
 \qquad \ell=L/2+2x_0.
\end{equation}
with $x_0=(3V-U)/(4V)$. The normalized discrete ground-state amplitudes have the large-$L$ form
\begin{equation}
    \psi_{m+1}\simeq\sqrt{\frac2\ell}\sin(\pi u_m)
 =\sqrt{\Delta u}\,\phi(u_m),
\end{equation}
with $\phi(u)=\sqrt2\sin(\pi u)$. Thus $\phi(u)$ is the continuum wavefunction, normalized by $\int_0^1|\phi(u)|^2\,du=1$.  The factor $\sqrt{\Delta u}$ is essential: $|\psi_{m+1}|^2$ is a probability for one discrete wall position, whereas $|\phi(u)|^2$ is a probability density, so that $|\psi_{m+1}|^2\simeq|\phi(u_m)|^2\Delta u$.

By the continuum limit we mean $L\to\infty$, the spacing $\Delta u$ tends to zero and the rescaled wall positions become dense in $[0,1]$. Therefore sums over wall
positions then become Riemann integrals.  For a cell with
$s=\lim_{L\to\infty}2j/L$, the condition $m\ge j$ becomes $u\ge s$. Consequently $P_j\to P(s)$, with
\begin{equation}
  P(s)=\int_s^1 2\sin^2(\pi u)\,du=1-s+\frac{\sin2\pi s}{2\pi},
  \label{eq:lv_profile}
\end{equation}
and the staggered density is $D=2P-1=1-2s+\sin(2\pi s)/\pi$.  The state is a coherent superposition of CDW configurations with the wall at different positions.  

To compute the QRE, observe that a one-body operator cannot connect distinct kink configurations, which differ by transferring at least two bosons.  Its off-diagonal expectation in any individual Fock configuration also vanishes.  Thus $C$ is diagonal, with the two
natural occupations $2P_j$ and $2-2P_j$ in cell $j$.  Defining $\Phi(P)=g(2P)+g(2-2P)$ gives
\begin{equation}
\Q^{(0)}=\sum_{j=1}^{L/2}\Phi(P_j).
\end{equation}
This finite sum is the full QRE of the leading wall state at size $L$; it is not yet the thermodynamic density $q_{\rm dw}$.  The latter is obtained only after dividing by $L$ and taking $L\to\infty$:
\begin{equation}
    q_{\rm dw}=\lim_{L\to\infty}\frac{\Q^{(0)}(L)}L
 =\lim_{L\to\infty}\frac1L\sum_{j=1}^{L/2}\Phi(P_j).
\end{equation}
Replacing this cell average by its continuum integral yields
\begin{equation}
  \begin{aligned}
  q_{\rm dw}&=\frac12\int_0^1\!ds\,\big[g(2P)+g(2-2P)\big]\\
  &=\int_0^1\!ds\,g\big(1+D(s)\big)=1.168491,
  \end{aligned}
  \label{eq:lv_qdw}
\end{equation}
using $P(1-s)=1-P(s)$.  

The limit used in Eq.~\eqref{eq:lv_qdw} discards the finite-size information. If $\Q^{(0)}=Lq_{\rm dw}+c_0^{(0)}+a^{(0)}/L+\cdots$, then
\begin{equation}
    \frac{\Q^{(0)}(L)}L
 =q_{\rm dw}+\frac{c_0^{(0)}}L+\frac{a^{(0)}}{L^2}+\cdots
\end{equation}
and both corrections vanish as $L\to\infty$.  To determine them we therefore return to the original finite sum and evaluate it more accurately, rather than taking the limit at the outset.  In particular, $c_0^{(0)}$ is the finite remainder after subtracting $Lq_{\rm dw}$, and $a^{(0)}/L$ is the next term in that remainder.

The connection between the finite sum and the continuum profile is made explicit by writing $\delta P_j=P_j-P(s_j)$ and expanding
\begin{equation}
 \Q^{(0)}(L)=\sum_j\Phi(P(s_j))
 +\sum_j\Phi'(P(s_j))\,\delta P_j+\cdots.
\end{equation}
The first sum will give $Lq_{\rm dw}+c_0^{(0)}$ through the order needed here; the second will give $a^{(0)}/L$. We evaluate them in that order below.  Both come from the same leading wall state, not from additional powers of the boson hopping $J$.  We will use $\Phi'(P)=2\Lambda(D)$, where
\begin{equation}
 \Lambda(D)=g'(1+D)-g'(1-D)
 =\ln\frac{(2+D)(1-D)}{(1+D)(2-D)}.
\end{equation}
Two discretization effects enter: the discrete profile $P_j$ differs from $P(s)$, and the sum over $j$ differs from the integral.

Using the effective length $\ell$ defined above, place cell $j$ at $s_j=(j-\tfrac12+x_0)/\ell$. $P_j$ is a sum of $|\psi_x|^2$ over $x\ge j+1$, i.e.\ a midpoint sum of
$(2/\ell)\sin^2(\pi u)$ from $u=s_j$ to $1$, up to $O(\ell^{-3})$ from the upper end, where the summand vanishes quadratically. The midpoint rule gives
\begin{equation}
  P_j=P(s_j)+\frac{\pi}{12\ell^2}\sin(2\pi s_j)+O(\ell^{-3}),
  \label{eq:lv_Pkink}
\end{equation}
because the derivative of $\sin^2(\pi u)$ is $\pi\sin2\pi u$, which vanishes at $u=1$. The points $s_j$ are the midpoints of a grid on $[x_0/\ell,1-x_0/\ell]$.  Near $s=0$, $1-P\simeq(2\pi^2/3)s^3$, so
$\Phi(P(s))-g(2)=O(s^3\ln s)$ and its first derivative vanishes at both ends of the interval.  The midpoint sum therefore equals the integral up to $O(\ell^{-3}\ln\ell)$:
\begin{equation}
  \sum_j\Phi(P(s_j))=2\ell q_{\rm dw}-2x_0g(2)+O(\ell^{-3}\ln\ell),
\end{equation}
The subtraction $2x_0g(2)$ comes from the two omitted endpoint intervals, each of width $x_0/\ell$, where $\Phi$ approaches $g(2)$. Since $2\ell=L+4x_0$, this expression already gives $Lq_{\rm dw}+2x_0[2q_{\rm dw}-g(2)]$: the second term is $c_0^{(0)}$.

The correction to the discrete probability profile supplies the next term.  Expanding $\Phi(P_j)$ to first order in $P_j-P(s_j)$ and using Eq.~\eqref{eq:lv_Pkink} gives
\begin{equation}
  \sum_j\frac{2\pi\Lambda(D_j)\sin2\pi s_j}{12\ell^2}
  \to\frac{\pi}{6\ell}\int_0^1\!ds\,\Lambda(D)\sin2\pi s .
\end{equation}
Each cell contributes an amount of order $\ell^{-2}$; summing over order $\ell$ cells therefore produces a correction of order $1/L$. Using $1/\ell=2/L+O(L^{-2})$ and combining both contributions,
\begin{equation}
  \begin{aligned}
  \Q^{(0)}&=Lq_{\rm dw}+c_0^{(0)}+\frac{a^{(0)}}{L}+O(L^{-2}),\\
  c_0^{(0)}&=2x_0\big[2q_{\rm dw}-g(2)\big],\\
  a^{(0)}&=\frac\pi3\int_0^1\!ds\,\Lambda\big(D(s)\big)\sin(2\pi s)=-1.24796 .
  \end{aligned}
  \label{eq:lv_c0a}
\end{equation}
Here $2q_{\rm dw}-g(2)=0.427439$, so $c_0^{(0)}=0.21372\,(3-U/V)$ for the kink.  The constant term says that the smeared profile extends over $2x_0$ more cells than the chain, each carrying $2q_{\rm dw}$ instead of $g(2)$. At this order the $1/L$ term comes only from the $\sin(2\pi s)/\ell^2$ correction of the discrete profile and does not depend on $x_0$.

\subsection{Corrections to the free-wall QRE density}
To compute the hopping correction to $q_{\rm dw}$, we first determine how virtual hops modify the one-body density matrix within a CDW domain. At $J=0$, choose the bulk CDW pattern $|2,0,2,0,\ldots\rangle$, with two bosons on each odd site (sublattice $A$) and zero on each even site (sublattice $B$).  A translation by one site gives the other CDW pattern, $|0,2,0,2,\ldots\rangle$.

To obtain the first-order correction, consider a wall configuration $|x\rangle$ with this CDW pattern around site $a\in A$.  A hop to the right ($\sigma=+1$) or left ($\sigma=-1$) produces the normalized Fock state $|e\rangle=b_{a+\sigma}^\dagger b_a|x\rangle/\sqrt{2}$.  The hop changes the occupations of the two sites from $(2,0)$ to $(1,1)$, so $|e\rangle$ is orthogonal to every wall configuration in $P$.  
Thus $P|e\rangle=0$ and $Q|e\rangle=|e\rangle$. These one-hop states are the eigenvectors of $Q$ with eigenvalue one that couple to $|x\rangle$ through $T$.  They also diagonalize $H_0$, which fixes the energy denominator in the first-order amplitude.  For a bulk hop the onsite energy decreases by $U$ and the nearest-neighbour interaction energy increases by $3V$, giving $E_e-E_0=\Delta_1=3V-U$.  The amplitude is therefore $\langle e|T|x\rangle/(E_0-E_e)=(-\sqrt{2}J)/(-\Delta_1)$.

Writing $|a\to a+\sigma\rangle$ for the corresponding one-hop state
of the bulk CDW reference, we obtain
\begin{equation}
  |\psi\rangle=|{\rm CDW}\rangle+\frac{\sqrt2J}{\Delta_1}
  \sum_{a\in A,\,\sigma=\pm1}|a\to a+\sigma\rangle+O(J^2).
\end{equation}
The same states enter the second-order energy as intermediate states
when a second hop returns the boson to its original site.  Summing over
the two hops available to each doubly occupied site gives the bulk
ground-state energy density $U/2-2J^2/\Delta_1$ through second order.

The first-order state gives nearest-neighbour coherence of order $J$. 
The second-order components that matter are the correlated two-hop states: the pair move `$0\,2$' (cost $4V$), the split `$1\,0\,1$' and the doubly occupied `$1\,2\,1$' ($4V-U$), two neighbours hopping the same way
`$1\,1\,1\,1\,2$' ($5V-2U$), and the onward transfer `$1\,0\,3$' ($U$). To order $J^2$,
\begin{equation}
  \begin{gathered}
  C_{AA}=2-t^2,\quad C_{BB}=t^2,\quad C_{i,i\pm1}=t,\\
  C_{B,B\pm2}=\tfrac{t^2}{2},\quad
  C_{A,A\pm2}=s_A=t^2+\frac{12J^2}{U\Delta_1}.
  \end{gathered}
  \label{eq:lv_Cbulk}
\end{equation}
Here $C_{AA}$ and $C_{BB}$ denote the mean occupations on the two sublattices. In $C_{A,A\pm2}$ and $C_{B,B\pm2}$, the displacement
counts physical sites: these are coherences between neighbouring sites
of the same sublattice.  The entry $C_{i,i\pm1}$ connects opposite
sublattices.  The transfer channel, with denominator $U$, enters only $s_A$.

For a free wall, the CDW orientation in a given cell depends on which side of the cell the wall occupies.  The two orientations are related by exchanging $A$ and $B$.  Away from the wall and the ends, their contribution to the one-body matrix through order $J^2$ is therefore obtained by weighting Eq.~\eqref{eq:lv_Cbulk} and its translated counterpart by $P_i$ and $1-P_i$, respectively. On an odd site,
\begin{equation}
  \begin{aligned}
  C_{ii}&=1+D_i-t^2D_i,\qquad C_{i,i\pm1}=t,\\
  C_{i,i\pm2}&=P_is_A+(1-P_i)\tfrac{t^2}{2},
  \end{aligned}
  \label{eq:lv_Cmix}
\end{equation}
and the same with $D\to-D$, $P\to1-P$ on an even site.  We checked this
element by element against ED at $L=12$.  The wall also produces
coherences $(2J/\Delta_2)(\psi_x+\psi_{x+1})^2$ on the bond where it hops
($2J/\Delta_2$ from each of the two configurations that share the
intermediate `$1\,1$').  They are $O(J/L)$ per bond and do not affect
$q_\infty$.

Using $w=s_A+t^2/2$, we now obtain the local natural occupations.
Since the matrix varies on the scale $L$, its spectrum follows from the
local Bloch bands (Weyl's law, exact for $q_\infty$).  For the two-site
cell at position $s$ and momentum $k$, the $2\times2$ symbol of
Eq.~\eqref{eq:lv_Cmix} has mean diagonal $1+w\cos k$, half-difference
$D\,X$ with $X=1-t^2+(s_A-\tfrac{t^2}2)\cos k$, and off-diagonal
$t(1+e^{-ik})$.  Hence
\begin{equation}
  \begin{gathered}
  n_\pm(s,k)=1+w\cos k\pm R,\\
  R^2=D^2X^2+2t^2(1+\cos k),
  \end{gathered}
  \label{eq:lv_bands}
\end{equation}
and
\begin{equation}
  q_\infty=\int_0^1\!ds\int_{-\pi}^{\pi}\frac{dk}{2\pi}\,
  \frac12\big[g(n_+)+g(n_-)\big].
  \label{eq:lv_lda}
\end{equation}
\textit{Order $J^2$.}  $F(m,R)=g(m+R)+g(m-R)$ is even in $R$, so it is an
analytic function of $m$ and $R^2$ wherever $m\pm R$ stay away from $0$.
$R^2$ is a polynomial in $D$, $t^2$ and $s_A$.  The expansion in $t^2$ and
$s_A$ is therefore regular near $D=0$, where $R\to0$ and the two
occupations cross: the apparent $1/|D|$ terms in an expansion of $R$
itself cancel between $g(n_+)$ and $g(n_-)$.  To first order,
\begin{equation}
  \delta F=\partial_mF\,w\cos k
  +\frac{\partial_RF}{2R}\,\delta(R^2),
\end{equation}
with $\delta(R^2)=2D^2[-t^2+(s_A-\frac{t^2}{2})\cos k]+2t^2(1+\cos k)$ and
$\partial_RF/2R=\Lambda(D)/2D$ at $m=1$, $R=|D|$.  All $\cos k$ terms
average to zero, including every term that contains $s_A$, and
\begin{equation}
  \kappa_2=\frac12\int_0^1\!ds\,\frac{1-D^2}{D}\,\Lambda(D)=-0.267288 .
  \label{eq:lv_kappa2}
\end{equation}
The integrand is regular, since $\Lambda(D)/D\to-1$ at $D=0$. 

The second-order terms are quadratic in $w$,
$t^2$ and $s_A-w=-t^2/2$.  They are not equally large:
$w/t^2=3(6V-U)/2U$, which is large when $V\gg U$, so the $w^2$ term
dominates, followed by $t^2w$ and $t^4$, smaller by $U/\Delta_1$ and
$(U/\Delta_1)^2$.  At fixed $U$ and large $V$, $w^2/t^2\to36J^2/U^2$: the
$w^2$ term is of the same order in $1/V$ as the $t^2$ term.  To obtain it
we set $t=0$, so that $s_A=w$ and
\begin{equation}
  n_\pm(s,k)=(1\pm D)(1+w\cos k).
\end{equation}
Expanding $g$ to second order and averaging over $k$,
$\langle g(n(1+w\cos k))\rangle_k=g(n)+\frac{w^2}{4}n^2g''(n)+O(w^4)$, with
$n^2g''(n)=-n/(n+1)$.  Summing the two bands,
\begin{equation}
  \begin{gathered}
  q_\infty=q_{\rm dw}+\kappa_2t^2-\kappa_ww^2+O(t^2w,t^4),\\
  \kappa_w=\frac18\int_0^1\!ds\,\frac{4-2D^2}{4-D^2}=0.101889 .
  \end{gathered}
  \label{eq:lv_kappaw}
\end{equation}

\subsection{From the QRE expansion to the finite-size estimators}
It is useful to distinguish the coefficient $b$ of a logarithm from the finite-size quantity $\beta_Q$.  Substitution of $\Q=Lq_\infty+b\ln L+c_0+a/L+e/L^2+\cdots$ gives
\begin{align}
 \beta_Q&=b-\frac{3a}{L\ln2}-\frac{21e}{L^2\ln2}+\cdots,\\
 q_{\rm eff}&=q_\infty
                 +\frac{4a}{L^2}+\frac{36e}{L^3}+\cdots.
\end{align}
Both combinations cancel $c_0$; the first also cancels the extensive term.  Thus $b=0$ does not imply $\beta_Q(L)=0$: the kink result $a^{(0)}=-1.24796$ gives $\beta_Q^{(0)}\simeq5.401/L$. These are leading-order
wall predictions; finite hopping also corrects $c_0$ and $a$. The absence of a leading logarithm follows here from the explicit profile sum, rather than from the bulk gap alone.

\section*{QRE in the Mott insulator at large \texorpdfstring{$U$}{U}}
\label{app:lu}
We consider Eq.~\eqref{eq:hamiltonian} at $V=0$, unit filling $N=L$, and $J/U\ll1$, with open boundaries.  At $J=0$ the unique ground state is the Mott product state $|\mathrm{MI}\rangle=|1\rangle^{\otimes L}$, whose one-body correlation matrix is $C=I$ and whose QRE is $\Q=Lg(1)=2L\ln2$.

A single hop on a bond creates either of the configurations
$|2,0\rangle$ and $|0,2\rangle$, each at excitation energy $U$.
The hopping matrix element is $-\sqrt{2}J$, giving a first-order
amplitude $\sqrt{2}J/U$ for each virtual doublon--holon configuration.
Interference with the Mott state therefore yields
\begin{equation}
 C_{i,i+1}=\langle b_i^\dagger b_{i+1}\rangle
 =4\frac JU+O\!\left((J/U)^3\right).
 \label{eq:mott_nn}
\end{equation}
The diagonal elements have no first-order correction, and correlations
beyond nearest neighbours start at second order.  Thus, writing
$\epsilon=J/U$,
\begin{equation}
 C=I+4\epsilon A+O(\epsilon^2),\qquad
 A_{ij}=\delta_{i,j+1}+\delta_{i,j-1},
 \label{eq:mott_C}
\end{equation}
where $A$ is the adjacency matrix of the open chain.

For the pure ground state, Eq.~\eqref{eq:gaussian_entropy} gives
$\Q=\Tr g(C)$.  Let $\delta C=C-I$.  Fixed unit filling implies
$\Tr\delta C=N-L=0$ exactly.  Expanding about $C=I$, with
$g'(1)=\ln2$ and $g''(1)=-1/2$, gives
\begin{align}
 \Q&=Lg(1)+g'(1)\Tr\delta C
       +\frac{g''(1)}2\Tr[(\delta C)^2]+\cdots\nonumber\\
     &=2L\ln2-\frac14\Tr[(\delta C)^2]+\cdots.
 \label{eq:mott_Q_expansion}
\end{align}
Since $\Tr A^2=2(L-1)$, the leading correction is
\begin{equation}
 \Q(L,0)=2L\ln2-8(L-1)\left(\frac JU\right)^2
              +O\!\left(L(J/U)^4\right).
 \label{eq:mott_Q_open}
\end{equation}
Odd powers vanish by the staggered gauge transformation $b_j\mapsto(-1)^j b_j$, which maps $J\mapsto-J$ and transforms $C$ by unitary conjugation.  Its spectrum, and hence $\Q$, is unchanged, so the analytic strong-coupling expansion is even in $J$.

Dividing Eq.~\eqref{eq:mott_Q_open} by $L$ and taking the thermodynamic
limit yields
\begin{equation}
 q_\infty(V=0)=2\ln2-8\left(\frac JU\right)^2
              +O\!\left((J/U)^4\right).
 \label{eq:mott_qinf}
\end{equation}
The factor $L-1$ in Eq.~\eqref{eq:mott_Q_open} retains the leading boundary correction from the missing bond. The negative bulk correction reflects the concavity of $g$: hopping broadens the natural occupation spectrum while preserving its mean.

\section*{Numerical methods and Mott--Haldane extrapolation}

\subsection{Numerical setup and occupation cutoff}
\label{app:numerics}
We compute ground states of open chains at unit filling, $N=L$, using DMRG~\cite{White1992,Schollwock2011}, with $J=1$ throughout. For the $V=0$ scan, we vary $U$ and consider chains of up to $L=4096$ sites. For the scans in $V$ at fixed $U/J=4$, the chain lengths range from $64$ to $512$ sites. The local bosonic basis is truncated to occupations $n=0,\ldots,n_{\max}$, with $n_{\max}=5$ for the production calculations. Checks at representative points in all phases using $n_{\max}=7$ yield no discernible changes in the reported observables. We use a maximum bond dimension of $2400$ and a singular-value cutoff of $10^{-8}$. The bond dimensions reached during the calculations remain below this cap. We assess the local occupation truncation by comparing calculations with $n_{\max}=4$ and $5$ at matching chain lengths and couplings. The largest absolute changes in the QRE density $\Q/L$ are $3.15\times10^{-5}$ for $2.30\leq V/J\leq2.60$ and $6.68\times10^{-5}$ for $3.27\leq V/J\leq3.30$. Both are below $7\times10^{-5}$ in the corresponding analysis windows. For the separate large-$V$ dataset in Fig.~\ref{fig:lv_scan}, we used $n_{\max}=4$, a maximum bond dimension of $128$, and a singular-value cutoff of $10^{-12}$. The $L=32$ results enter only the estimator $\beta_Q(128)$. 

\subsection{Correlation functions and order parameters}
For each ground state, the one-body correlation matrix
$C_{ij}=\langle c_i^\dagger c_j\rangle$ gives the natural occupations
used to compute the QRE through Eq.~\eqref{eq:gaussian_entropy}.
The conventional order parameters are evaluated on the same ground
states. Writing $\delta n_i=n_i-1$, we use the string and staggered
density correlators
\begin{align*}
 C_{\mathrm s}(i,j)&=\left\langle\delta n_i
 e^{i\pi\sum_{i<k<j}\delta n_k}\delta n_j\right\rangle,\\
 C_{\mathrm{CDW}}(i,j)&=(-1)^{j-i}\langle\delta n_i\delta n_j\rangle.
\end{align*}
Their long-distance limits define string and density-wave order,
taking the magnitude for string order. For the finite-chain
comparison, $O_{\mathrm s}(L,V)=|C_{\mathrm s}(i,j)|$ uses endpoints
centered in the chain at separation $j-i=L/2$.
This fixes the string observable used in the exponent estimates
and extrapolations.

\subsection{Mott--Haldane peak and string-order crossing}
\label{app:stability}
We compare markers obtained from the same saved ground states, with
occupation cutoffs $n_{\max}=4,5$ and coupling spacing $\Delta(V/J)=0.005$.
The five QRE triples $(L/4,L/2,L)$ use thirteen distinct physical lengths.
Writing $v=V/J$, define $v_{\mathrm p}(L)$ by the maximum of
$\beta_Q(L,v)$ and the string marker $v_{\mathrm s}(L)$ by the zero of
\begin{equation}
 R_{\mathrm s}(L,v)=\frac{1}{\ln2}
 \ln\!\left[\frac{O_{\mathrm s}(L/4,v)O_{\mathrm s}(L,v)}
 {O_{\mathrm s}(L/2,v)^2}\right].
 \label{eq:app_string_marker}
\end{equation}

\begin{figure}[h!]
 \centering
 \includegraphics[width=\columnwidth]{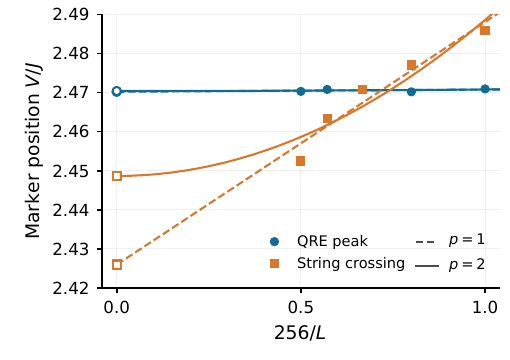}
 \caption{MI--HI marker stability from the same ground states.
 QRE peak positions and string-order crossings for largest lengths
 $L=256,320,384,448,512$. Lines fit Eq.~\eqref{eq:app_peak_drift}
 with $p=1$ (dashed) and $p=2$ (solid); open symbols mark extrapolated
 intercepts.}
 \label{fig:app_mihi}
\end{figure}

This compares effective string exponents on successive size pairs; it vanishes for a size-independent power-law exponent. Three-point quadratic interpolation gives QRE peaks between $2.47019$ and $2.47094$. In contrast, the string crossings range from $2.45247$ to $2.48579$ [Fig.~\ref{fig:app_mihi}]. 
The QRE marker therefore varies about $45$ times less across these lengths. This quantifies reduced finite-size drift, rather than an equivalent gain in statistical precision for the thermodynamic critical point.

To examine extrapolation, we fit either marker to
\begin{equation}
 v_m(L)=v_m^{\infty}+A_m(256/L)^p,
 \qquad m\in\{\mathrm p,\mathrm s\}.
 \label{eq:app_peak_drift}
\end{equation}
For the QRE peak, the all-size $p=1$ and $p=2$ fits give
$v_{\mathrm p}^{\infty}=\QREFitOne$ and $\QREFitTwo$, respectively.
Applying the same fits to string order gives $\StringFitOne$ and $\StringFitTwo$ [Fig.~\ref{fig:app_mihi}]. The QRE estimate is therefore much less sensitive to the choice between these drift models. For string order, we also include $p=0.60,0.66$, both occupation cutoffs, and omission of the smallest one or two sizes or each size in turn. Their spread, rounded to $V/J=\StringSpreadLow$--$\StringSpreadHigh$, defines the orange band in Figs.~\ref{fig:qre_transitions} (a,b); it measures tested fit sensitivity, not the total uncertainty in the thermodynamic transition.

The QRE peak remains stable under local quadratic, cubic, and quartic fits to three to nine points: all cutoff-five estimates lie in $2.47015$--$2.47107$. Coarsening the grid by factors two and three gives a comparable range. Changing the occupation cutoff shifts a three-point peak by at most $5.7\times10^{-4}$. These checks support
localization near $\MIQRECenter$. 

\label{page:appendix-end}

\onecolumngrid
\end{document}